\documentclass[fullpage, a4]{article}

\iftrue		
	\usepackage[dvips]{epsfig}
	\def\grafico#1{\epsfig{figure=#1.eps}}
\else		
	\usepackage{graphicx}
	\usepackage{epstopdf}
	\def\grafico#1{\includegraphics[width=45ex,height=30ex]{#1}}
\fi

\begin{document}

\title{Solutions of all one-dimensional wave equations with time independent
potential and separable variables}

\author{{Marco Ferraris$^a$, Alessandro D.A.M. Spallicci$^b$\footnote{Corresponding author. Email: spallicci@obs-nice.fr. Postal address: Bd. de 
l'Observatoire, BP 4229, 06304 Nice. Tel. +33492003111, Cell. + 33623153584, Fax + 334920031338}}\\
{$^a$ Ist. Fisica Matematica J.-L. Lagrange,}\\
{Dip. Matematica, Univ. di Torino}\\
{$^b$ ARTEMIS, D\'ep. d'Astrophysique Relativiste,}\\
{Obs. de la C\^ote d'Azur, Nice}}
\date{21 January 2004}

\maketitle
\begin{abstract}
Solutions, in terms of special functions, of all wave equations
$u_{xx} - u_{tt} = V(x) u(t,x)$, characterised by eight inequivalent time
independent potentials and by variables separation, have been found. 
The real valueness and  the properties 
of the solutions produced by computer algebra programs are not always
manifest and in this work we provide ready to use solutions. We discuss
especially the potential $(m_1 + m_2 \sinh x)\cosh^{-2}x$. Such potential
approximates the Schwarzschild black hole potential for even parity and its use  
for determining black holes quasi-normal modes is hinted to.
\end{abstract}


\section{Introduction}

In the frame of our research work on analytic solutions of black holes
differential equations \cite{Spallicci}, we have determined the analytic
solutions of all wave equations with time independent potential $V(x)$:
\begin{eqnarray}
&&mx  \label{eq:1p} \\
&&mx^{-2}  \label{eq:2p} \\
&&m\sin ^{-2}x  \label{eq:3p} \\
&&m\sinh ^{-2}x  \label{eq:4p} \\
&&m\cosh ^{-2}x  \label{eq:5p} \\
&&m\exp x  \label{eq:6p} \\
&&(m_{1}+m_{2}\sin x)\cos ^{-2}x  \label{eq:7p} \\
&&(m_{1}+m_{2}\sinh x)\cosh ^{-2}x  \label{eq:8p} \\
&&(m_{1}+m_{2}\cosh x)\sinh ^{-2}x  \label{eq:9p} \\
&&m_{1}e^{x}+m_{2}e^{2x}  \label{eq:10p} \\
&&m_{1}+m_{2}x^{-2}  \label{eq:11p} \\
&&m  \label{eq:12p}
\end{eqnarray}
for the wave equation:
\begin{equation}
u_{xx}-u_{tt}=V(x)u(t,x)  \label{eq:allw}
\end{equation}
These potentials characterize the wave equation by variable separation \cite
{Zhdanov}. They may be reduced to eight irreducible forms. The potential (%
\ref{eq:6p}) is equivalent to potential (\ref{eq:12p}) with the change of
variables:
\[
x^{\prime }=\exp \left( \frac{x}{2}\right) \cosh \frac{t}{2}\ \ \ \ \ \ \ \
\ \ t^{\prime }=\exp \left( \frac{x}{2}\right) \sinh \frac{t}{2}
\]
while potentials (\ref{eq:3p},\ref{eq:4p},\ref{eq:5p}) are equivalent to
potential (\ref{eq:2p}) with the following change of variables respectively:
\begin{eqnarray}
x^{\prime }=\tan \xi +\tan \eta  &\ \ \ \ \ \ \ \ \ \ &t^{\prime }=\tan \xi
-\tan \eta  \\
x^{\prime }=\tanh \xi +\tanh \eta  &\ \ \ \ \ \ \ \ \ \ &t^{\prime }=\tanh
\xi -\tanh \eta  \\
x^{\prime }=\coth \xi +\tanh \eta  &\ \ \ \ \ \ \ \ \ \ &t^{\prime }=\coth
\xi -\tanh \eta 
\end{eqnarray}
where
\[
\xi =\frac{1}{2}(x+t)\ \ \ \ \ \ \ \ \ \ \ \ \ \ \ \ \ \ \ \ \eta =\frac{1}{2%
}(x-t)
\]
The general form of the solution with separated variables of eq.(\ref
{eq:allw}) is:
\begin{equation}
u(t,x)=\phi _{1}(\omega _{1})\phi _{2}(\omega _{2})
\end{equation}
where $\omega _{1}=\omega _{1}(t,x)$, $\omega _{2}=\omega _{2}(t,x)$, and $%
\phi _{1}(\omega _{1})$ and $\phi _{2}(\omega _{2})$ are arbitrary solutions
of the separated ordinary differential equation:
\begin{equation}
\frac{d^{2}\phi _{i}}{d\omega _{i}^{2}}=[(c_{i}+g_{i}(\omega _{i})]\phi _{i}
\end{equation}
$c_{i}$ being the separation constant and $i=t,x$.

The wave equations separate in several coordinate systems, among which the
commonest and simplest poses $\omega _{1}=t$, $\omega _{2}=x$, $g_{t}=0$ and 
$g_{x}=V(x)$. With the support of Maple 7 software, we have obtained the
general solutions of all differential equations, analysed their properties
and proved their real valueness. The solution for the time dependent
function $\phi _{t}$ equation:
\begin{equation}
\frac{d^{2}\phi _{t}}{dt^{2}}-c_{t}\phi _{t}=0
\end{equation}
is:
\begin{equation}
\phi _{t}=k_{1}e^{i\sqrt{c_{t}}t}+k_{2}e^{-i\sqrt{c_{t}}t}
\end{equation}
The space dependent solutions in the appendix have been obtained by
simplifying the results returned by the \emph{odesolve} function of Maple 7.
The solutions corresponding to potentials (\ref{eq:7p},\ref{eq:8p},\ref
{eq:9p}) require an additional effort to obtain a readable form\footnote{%
Ref. \cite{Abramowitz} has been used for special function properties.}. It
is the latter we directly show in the appendix. The applicability, the
properties, and the real valueness of the solution corresponding to
potential (\ref{eq:8p}) have been considered in detail in the next section.

\section{The Regge--Wheeler--Zerilli equation}

Regge and Wheeler \cite{ReggeWheeler} proved the stability of the
Schwarzschild black hole in vacuum for axial perturbations, while Zerilli 
\cite{Zerilli} found the equation for polar perturbations. The latter is
written in terms of the wave function $\Psi _{l}$, for each $l$--pole
component :
\begin{equation}
\frac{d^{2}\Psi _{l}(r,t)}{dr^{*2}}-\frac{d^{2}\Psi _{l}(r,t)}{dt^{2}}%
-V_{l}(r)\Psi _{l}(r,t)=0  \label{eq:rwz*}
\end{equation}
where
\begin{equation}
r^{*}=r+2M\ln \left( {{\frac{r}{2M}}}-1\right)   \label{eq:rstarr}
\end{equation}
is the tortoise coordinate and the potential $V_{l}(r)$ is:
\begin{equation}
V_{l}(r)=\left( 1-\frac{2M}{r}\right) \frac{2\lambda ^{2}(\lambda
+1)r^{3}+6\lambda ^{2}Mr^{2}+18\lambda M^{2}r+18M^{3}}{r^{3}(\lambda
r+3M)^{2}}  \label{eq:potz}
\end{equation}
while $\lambda ={\frac{1}{2}}(l-1)(l+2)$. The Zerilli potential (\ref
{eq:potz}) may be approximated in a selected radial coordinate domain,
including the maximum, by the potential (\ref{eq:8p}). Blome and Mashhoon%
\cite{blma84} have used the Eckart potential \cite{ec30}:
\begin{equation}
V=V_{0}e^{2\mu }-V_{0}\{\tanh [\alpha (x-x_{0})+\mu ]-\tanh \mu \}^{2}\cosh
^{2}\mu 
\end{equation}
while Ferrari and Mashhoon\cite{fema84}, Beyer\cite{be99} have used the
P\"{o}schl--Teller potential\cite{pote33}:
\begin{equation}
V=\frac{V_{0}}{\cosh ^{2}\alpha (x-x_{0})}
\end{equation}
for derivation of the QNM (quasi--normal modes) of a black hole. The ground
state plus the first few excited states can be approximated by the bound
states of the inverted potential. We note that (\ref{eq:8p}) well reproduces
the Zerilli potential and investigations on quasi--normal modes are
undergoing\cite{aofesp03}. The Zerilli potential has not allowed any
analytic determination of the QNM\footnote{%
Black holes perturbations equations do not admit exact solutions, apart of
approximate solutions for portions of the frequency domain, e.g. \cite{ip71}--%
\cite{fi02}, or post--Newtonian expansions in the weak field and slow motion,
e.g. \cite{sa94}.}. The polar potential is thus substituted by:
\begin{equation}
V_{l}^{\prime }(x)=A[m_{1}+m_{2}\sinh (kr^{*})]\cosh ^{-2}(kr^{*})
\end{equation}
where $A,m_{1},m_{2},k$ are parameters depending on $l$ and $M$, for proper
curve fitting.

\begin{figure}[tbp]
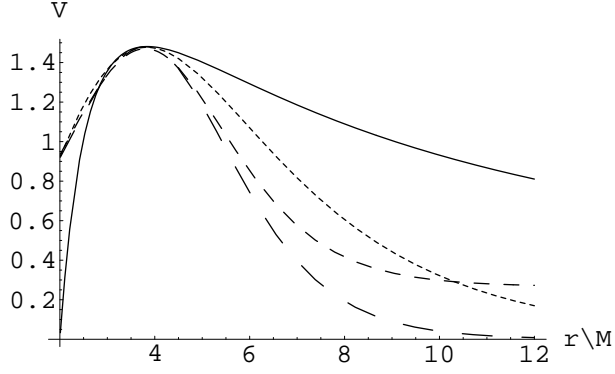

\begin{center}
\grafico{similpot}
\end{center}
\caption{The Zerilli potential, the potential (\ref{eq:8p}), the Eckart and the
P\"{o}schl--Teller potentials for $l=2$ (quadrupole), starting from above. On
the $y$ axis the potential in ($1/M^{2}$) units; on the $x$ axis the radial
coordinate in $r/M$ units. }
\label{fig:similpot}
\end{figure}

From \textit{fig. 1}, we evince that the original black hole Zerilli
potential is best replaced by potential (\ref{eq:8p}) for $r/M<10$ and after by the
Eckart potential.

The general form of the solution is:
\begin{equation}
\Psi _{l}(r^{*},t)=\psi _{lr^{*}}[\omega _{r^{*}}(r^{*},t)]\psi _{lt}[\omega
_{t}(r^{*},t)]
\end{equation}
where $\psi _{li},i=r^{*},t$ are solutions of the o.d.e.:
\begin{equation}
\frac{d^{2}\psi _{li}(\omega _{i})}{d\omega _{i}^{2}}=[c_{i}+g_{i}(\omega
_{i})]\psi _{li}(\omega _{i})  \label{eq:zhd}
\end{equation}
The d'Alembert equation $[\partial _{t}^{2}-\partial _{r^{*}}^{2}-V^{\prime
}(r^{*})]\Psi _{l}=0$ separates into four coordinate systems, among which
the following is the only with an explicit relation for $\omega _{i}$:
\begin{equation}
\omega _{r^{*}}=r^{*}~~~~~\omega _{t}=t~~~~~f_{r^{*}}=A[m_{1}+m_{2}\sinh
(kr^{*})]\cosh ^{-2}(kr^{*})~~~~~f_{t}=0
\end{equation}
In order to write the solution corresponding to the potential (\ref{eq:8p})
in a more suitable way, we assume $c_{x}<0$ and we introduce new real
parameters $k$, $x$, $\mu _{1}$, $\mu _{2}$ and $\sigma $ such that:
\[
m_{1}=1/4-\mu _{1}^{2}+\mu _{2}^{2},~~m_{2}=2\mu _{1}\mu
_{2},~~A=k^{2},~~c_{x}=-\sigma ^{2}k^{2},~~\sinh (kr^{*})=x
\]
Consequently:
\begin{eqnarray*}
\alpha _{3} &=&2(\mu _{1}+i\mu _{2}) \\
a &=&1/2+\mu _{1}-i\sigma  \\
b &=&1/2+\mu _{1}+i\sigma  \\
c &=&1+\mu _{1}+i\mu _{2} \\
z &=&(1-ix)/2
\end{eqnarray*}
Accordingly, the potential (\ref{eq:potz}) can be rewritten as follows:
\begin{eqnarray}
\psi _{lr^{*}} &=&(x-i)^{(\mu _{1}+1/2-i\mu _{2})/2}(x+i)^{(\mu
_{1}+1/2+i\mu _{2})/2}w(z)  \nonumber \\
&=&\sqrt{x^{2}+1}^{(\mu _{1}+1/2)}e^{-\mu _{2}\arctan (1/x)}w(z)
\end{eqnarray}
where the function $w(z)$ has to be a solution of the hypergeometric
equation: 
\begin{equation}
z(z-1)\frac{d^{2}w(z)}{dz^{2}}+[z(1+b+a)-c)]\frac{dw(z)}{dz}+abw(z)=0
\end{equation}
and the general solution of which is: 
\begin{equation}
w(z)=k_{1}F(a,b;c;z)+k_{2}z^{1-c}F(a-c+1,b-c+1;2-c;z)  \label{soluz}
\end{equation}
Eq.(\ref{eq:zhd}) has, therefore, the following expression:
\begin{equation}
\psi _{lr^{*}}=\left[ \cosh (kr^{*})\right] ^{(\mu _{1}+1/2)}e^{-\mu
_{2}\arctan [1/\sinh (kr^{*})]}w\left( \frac{1-i\sinh kr^{*}}{2}\right) 
\end{equation}
The solutions (\ref{soluz}) must be real valued functions. This is possible
since:
\begin{equation}
\bar{a}=b,~~~\bar{b}=a,~~~\bar{c}=a+b+1-c,~~~\bar{z}=1-z
\end{equation}
so that:
\begin{equation}
\overline{F(a,b;c;z)}=F(\bar{a},\bar{b};\bar{c};\bar{z})=F(b,a;a+b+1-c;1-z)
\end{equation}
and, moreover, for one of the identities between hypergeometric functions we
know that:
\begin{eqnarray}
F(b,a;a+b+1-c;1-z) &=&\sigma_1 F(a,b;c;z)  \\
&&+\sigma_2 z^{1-c}F(a-c+1,b-c+1;2-c;z)\nonumber
\end{eqnarray}
where we set:
\begin{equation}
\sigma_1 = \frac{{\Gamma (\bar{c})\Gamma (1-c)}}{{\Gamma
(a-c+1)\Gamma (b-c+1)}}, \ \ \ \ 
\sigma_2 = \frac{{\Gamma (\bar{c})\Gamma (1-c)}}{{\Gamma (a)\Gamma (b)}}
\end{equation}

Finally, for the potential (\ref{eq:8p}) the space dependent solution is:
\begin{eqnarray}
\phi _{x} &=&z^{(c/2-1/4)}(1-z)^{[(a+b-c)/2+1/4]}  \nonumber \\
&&\left[ k_{1}\hbox{F}(a,b;c;z)+k_{2}z^{(1-c)}\hbox{F}(1+a-c,1+b-c;2-c;z)%
\right] 
\end{eqnarray}
where
\begin{eqnarray*}
a &=&(\alpha _{3}+\alpha _{4})/4+\beta +1/2 \\
b &=&(\alpha _{3}+\alpha _{4})/4-\beta +1/2 \\
c &=&1+\alpha _{3}/2 \\
z &=&(1\pm i\sinh {x})/2
\end{eqnarray*}
with $\alpha _{3}=\pm \sqrt{1-4m_{1}\pm 4im_{2}}$, $\alpha _{4}=\pm \sqrt{%
1-4m_{1}\mp 4im_{2}}$ and $\beta =\pm \sqrt{c_{x}}$.

\section{Conclusions}

Solutions, in terms of special functions, of all wave equations $%
u_{xx}-u_{tt}=V(x)u$, characterized by eight inequivalent time independent
potentials and by variable separation, have been found. One among the
potentials, $(m_{1}+m_{2}\sinh x)\cosh ^{-2}x$, has a shape similar to
Schwarzschild black hole potential for polar perturbations.

\section{Acknowledgments}

AS acknowledges the European Space Agency for awarding him the Senior
Research Fellowship G. Colombo at the Observatoire de la C\^ote d'Azur, Nice.

\section{Appendix}

In this section we write the space dependent solutions for all potentials,
obtained after some computation. For the potential (\ref{eq:1p}), the
solution can be written in terms of the Airy functions:
\begin{equation}
\phi _{x}=k_{1}\hbox{Ai}\left( \frac{c_{x}+mx}{m^{2/3}}\right) +k_{2}%
\hbox{Bi}\left( \frac{c_{x}+mx}{m^{2/3}}\right) 
\end{equation}
or in terms of Bessel $\hbox{K}$ and $\hbox{I}$ functions:
\begin{eqnarray}
\phi _{x} &=&\sqrt{c_{x}+mx}\left\{ k_{1}\hbox{K}\left( \frac{1}{3},\frac{%
2(c_{x}+mx)^{3/2}}{3\left| m\right| }\right) \right.   \nonumber \\
&&\left. +k_{2}\left[ \hbox{I}\left( -\frac{1}{3},\frac{2(c_{x}+mx)^{3/2}}{%
3\left| m\right| }\right) +\hbox{I}\left( \frac{1}{3},\frac{2(c_{x}+mx)^{3/2}%
}{3\left| m\right| }\right) \right] \right\} 
\end{eqnarray}

For the potential (\ref{eq:2p}):
\begin{equation}
\phi _{x}=\sqrt{x}\left[ k_{1}\hbox{J}\left( \frac{\sqrt{1+4m}}{2},x\sqrt{%
-c_{x}}\right) +k_{2}\hbox{Y}\left( \frac{\sqrt{1+4m}}{2},x\sqrt{-c_{x}}%
\right) \right] 
\end{equation}
where $\hbox{J}$ and $\hbox{Y}$ are Bessel functions.

For the potential (\ref{eq:3p}):
\begin{eqnarray}
\phi _{x} &=&\sqrt{\sin x}\left[ k_{1}\hbox{P}\left( i\sqrt{c_{x}}-\frac{1}{2%
},\frac{1}{2}\sqrt{1+4m},\cos x\right) \right.   \nonumber \\
&&\left. +k_{2}\hbox{Q}\left( i\sqrt{c_{x}}-\frac{1}{2},\frac{1}{2}\sqrt{1+4m%
},\cos x\right) \right] 
\end{eqnarray}
where $\hbox{P}(\nu ,\mu ,z)=\hbox{P}_{\nu }^{\mu }(z)$ and $\hbox{Q}(\nu
,\mu ,z)=\hbox{Q}_{\nu }^{\mu }(z)$ are associated Legendre functions of the
first and second kind, respectively.

For the potential (\ref{eq:4p}):
\begin{eqnarray}
\phi _{x} &=&\sqrt{\sinh x}\left[ k_{1}\hbox{P}\left( \sqrt{c_{x}}-\frac{1}{2%
},\frac{1}{2}\sqrt{1+4m},\cosh x\right) \right.   \nonumber \\
&&\left. +k_{2}\hbox{Q}\left( \sqrt{c_{x}}-\frac{1}{2},\frac{1}{2}\sqrt{1+4m}%
,\cosh x\right) \right] 
\end{eqnarray}

For the potential (\ref{eq:5p}):
\begin{eqnarray}
\phi _{x} &=&\sqrt{\cosh x}\left[ k_{1}\hbox{P}\left( \sqrt{c_{x}}-\frac{1}{2%
},\frac{1}{2}\sqrt{1-4m},i\left| \sinh x\right| \right) \right.   \nonumber
\\
&&\left. +k_{2}\hbox{Q}\left( \sqrt{c_{x}}-\frac{1}{2},\frac{1}{2}\sqrt{1-4m}%
,i\left| \sinh x\right| \right) \right] 
\end{eqnarray}

For the potential (\ref{eq:6p}):
\begin{equation}
\phi _{x}=k_{1}\hbox{J}\left( 2\sqrt{c_{x}},2\sqrt{-m}~e^{x/2}\right) +k_{2}%
\hbox{Y}\left( 2\sqrt{c_{x}},2\sqrt{-m}~e^{x/2}\right) 
\end{equation}

For the potential (\ref{eq:7p}):
\begin{eqnarray}
\phi _{x} &=&z^{(c/2-1/4)}(1-z)^{[(a+b-c)/2+1/4]}  \nonumber \\
&&\left[ k_{1}\hbox{F}(a,b;c;z)+k_{2}z^{(1-c)}\hbox{F}(1+a-c,1+b-c;2-c;z)%
\right] 
\end{eqnarray}
where $\hbox{F}$ indicates Gauss hypergeometric function ${{}_{2}\hbox{F}_{1}%
}$ and the parameters $a$, $b$, $c$ and $z$ are defined by:
\begin{eqnarray*}
a &=&(\alpha _{1}+\alpha _{2})/4+\beta +1/2 \\
b &=&(\alpha _{1}+\alpha _{2})/4-\beta +1/2 \\
c &=&1+\alpha _{2}/2 \\
z &=&(1+\sin x)/2
\end{eqnarray*}
with $\alpha _{1}=\pm \sqrt{1+4m_{1}+4m_{2}}$, $\alpha _{2}=\pm \sqrt{%
1+4m_{1}-4m_{2}}$ and $\beta =\pm \sqrt{-c_{x}}$.

For the potential (\ref{eq:9p}):
\begin{eqnarray}
\phi _{x} &=&z^{(c/2-1/4)}(z-1)^{[(a+b-c)/2+1/4]}  \nonumber \\
&&\left[ k_{1}\hbox{F}(a,b;c;z)+k_{2}z^{(1-c)}\hbox{F}(1+b-c,1+a-c;2-c;z)%
\right] 
\end{eqnarray}
where
\begin{eqnarray*}
a &=&(\alpha _{5}+\alpha _{6})/4+\beta +1/2 \\
b &=&(\alpha _{5}+\alpha _{6})/4-\beta +1/2 \\
c &=&1+\alpha _{6}/2 \\
z &=&(1+\cosh x)/2
\end{eqnarray*}
with $\alpha _{5}=\pm \sqrt{1+m_{1}+m_{2}}$, $\alpha _{6}=\pm \sqrt{%
1+m_{1}-m_{2}}$ and $\beta =\pm \sqrt{c_{x}}$.

For the potential (\ref{eq:10p}):
\begin{equation}
\phi _{x}=e^{-x/2}\left[ k_{1}\hbox{M}\left( -\frac{m_{1}}{2\sqrt{m_{2}}}%
\sqrt{c_{x}},-2\sqrt{m_{2}}~e^{x}\right) +k_{2}\hbox{W}\left( -\frac{m_{1}}{2%
\sqrt{m_{2}}},\sqrt{c_{x}},-2\sqrt{m_{2}}~e^{x}\right) \right] 
\end{equation}
where $\hbox{W}$ and $\hbox{M}$ are Whittaker's functions.

For the potential (\ref{eq:11p}), the space dependent solution is equal to
the solution for the potential (\ref{eq:2p}), while for the potential (\ref
{eq:12p}), the solution is trivial:
\begin{equation}
\phi _{x}=K_{1}e^{\sqrt{c_{x}}x}+K_{2}e^{-\sqrt{c_{x}}x}
\end{equation}


\begin{thebibliography}{99}
\bibitem{Spallicci}  Spallicci A., 2000. \emph{Analytic solution of
Regge--Wheeler differential equation for black hole perturbations in radial
coordinate and time domains}, 13th Italian Conf. on General Rel. and Grav.
Physics, 21--25 September 1998 Monopoli (Bari), Springer, 371.

\bibitem{Zhdanov}  Zhdanov R.Z., Revenko I.V., Fushchych W.I., 1993. \emph{%
Orthogonal and non--orthogonal separation of variables in the wave equation }%
$u_{tt}-u_{xx}+V(x)=0$, J. Phys. A: Math. Gen., \textbf{26}, 5959.

\bibitem{Abramowitz}  Abramowitz M., Stegun I.A., 1965. \emph{Handbook of
Mathematical Functions}, Dover.

\bibitem{ReggeWheeler}  Regge T., Wheeler J.A, 1957. \textit{Stability of a
Schwarzschild singularity}, Phys. Rev., \textbf{108}, 1063.

\bibitem{Zerilli}  Zerilli F.J., 1970a. \emph{Effective potential for
even--parity Regge--Wheeler gravitational perturbation equations}, Phys.
Rev. D, \textbf{2}, 2141.

\bibitem{blma84}  Blome H.-J., Mashhoon B., 1984. \emph{Quasi--normal
oscillations of a Schwarzschild black hole}, Phys. Lett. A, \textbf{100},
231.

\bibitem{ec30}  Eckart C., 1930. \emph{The penetration of a potential
barrier by electrons}, Phys. Rev., \textbf{35}, 1303.

\bibitem{fema84}  Ferrari V., Mashhoon B., 1984. \emph{New approach to the
quasinormal modes of a black hole}, Phys. Rev. D, \textbf{30}, 1984.

\bibitem{be99}  Beyer H.R., 1999. \emph{On the completeness of the
quasinormal modes of the P\"{o}schl--Teller potential}, Comm. Math. Phys., 
\textbf{204}, 397.

\bibitem{pote33}  P\"{o}schl G., Teller E., 1933. \emph{Bemerkungen zur
quantenmechanik des anharmonischen oszillators}, Z. Phys., \textbf{83}. 143.

\bibitem{aofesp03}  Aoudia S., Ferraris M., Spallicci A. (in preparation).

\bibitem{ip71}  Ipser J.R., 1971. \emph{Gravitational radiation from slowly
rotating, fully relativistic stars}, AP. J., \textbf{166}, 175.

\bibitem{fa71}  Fackerell E.D., 1971. \emph{Solutions of Zerilli's equation
for even--parity gravitational perturbations.}, Ap. J., \textbf{166}, 197.

\bibitem{le85}  Leaver E.W., 1985. \emph{An analytic representation for the
quasi--normal modes of Kerr black holes}, Proc. Roy. Soc. Lond. A, \textbf{%
402}, 285.

\bibitem{le86a}  Leaver E.W., 1986a. \emph{Solutions to a generalized
spheroidal wave equation: Teukolsky's equations in general relativity, and
the two--center problem in molecular quantum mechanics} J. Math. Phys., 
\textbf{27}, 1238.

\bibitem{le86b}  Leaver E.W., 1986b. \emph{Spectral decomposition of the
spectral response of the Schwarzschild geometry}, Phys. Rev. D, \textbf{34}, 384.

\bibitem{masuta96}  Mano S., Suzuki H., Takasugi E., 1996. \emph{Analytic
solutions of the Regge--Wheeler equation and the post--Minkowskian expansion}%
, Progr. Theor. Phys., \textbf{96}, 549.

\bibitem{fi02}  Figueiredo B.D.B., 2002. \emph{On some solutions of
spheroidal wave equations and applications}, J. Phys. A., \textbf{35}, 2877.
Errata corrige J. Phys. A., \textbf{35}, 4799.

\bibitem{sa94}  Sasaki M., 1994. \emph{Post--Newtonian expansion of the
ingoing--wave Regge--Wheeler function}, Progr. Theor. Phys., \textbf{92}, 17.
\end{thebibliography}
\end{document}